\documentclass[conference]{IEEEtran}
\ifCLASSINFOpdf
\else
\fi

\usepackage{amsmath}
\usepackage{amsfonts}
\usepackage{amssymb}   
\usepackage{bm}
\usepackage[]{algorithm2e}
\usepackage{algpseudocode}
\usepackage{graphicx}
\newcommand{\norm}[1]{\left\lVert#1\right\rVert}
\usepackage{pgfplots} 
\usepackage[fleqn]{mathtools}
\pgfplotsset{compat=newest} 
\pgfplotsset{plot coordinates/math parser=false} 
\newlength\figureheight 
\newlength\figurewidth 
\allowdisplaybreaks[2]

\begin{document}
%
\title{Joint beamforming and network topology optimization of green cloud radio access networks}

\author{\IEEEauthorblockN{Alaa Alameer and Aydin Sezgin}
\IEEEauthorblockA{Ruhr-Universitat Bochum, 44780, Germany \\
Email: $\left\lbrace\text{alaa.alameerahmad, aydin} \right\rbrace$@rub.de}}


%


\maketitle
\vspace{-1em}
\begin{abstract}
Cloud radio access networks (C-RAN) are a promising technology to enable the ambitious vision of the fifth-generation (5G) communication networks. In spite of the potential benefits of C-RAN, the operational costs are still a challenging issue, mainly due to the centralized processing scheme and the large number of operating remote radio head (RRH) connecting to the cloud. In this work we consider a setup in which a C-RAN is powered partially with a set of renewable energy sources (RESs), our aim is to minimize the processing/backhauling costs at the cloud center as well as the transmission power at the RRHs, while satisfying some user quality of service (QoS). This problem is first formulated as a mixed integer non linear program (MINLP) with a large number of optimization variables. The underlying NLP is non-convex, though we address this issue through reformulating the problem using the mean squared error (MSE)-rate relation. To account to the large-scale of the problem, we introduce slack variables to decompose the reformulated (MINLP) and enable the application of a distributed optimization framework by using the alternating direction method of multipliers (ADMM) algorithm.
\end{abstract}

\IEEEpeerreviewmaketitle

\section{Introduction}
Future cellular networks are expected to face a drastic increase in data flow, thanks to the dramatic expansion of the number of devices connecting to the network. 
The performance of the (5G) networks is expected to go far beyond the limits offered by today's networks in terms of data rates and latency to account for such a tremendous demand for data.
C-RAN are a promising technology to realize 5G networks. 
In C-RAN, the functionality of conventional base stations is reduced to a simple transmission/reception point, while the signal processing and encoding/decoding tasks are moved to a centralized utility, referred to as cloud center.                    
Hence, through dense deployment of RRHs, C-RAN can achieve higher data rates and spectral efficiency compared to classical cellular networks [1].  
However, the joint baseband data processing and cooperation among large number of RRHs results in a huge traffic on backhaul links to the BBU. Thus, RRHs clustering, i.e, characterizing the minimum set of RRH able to satisfy some QoS at each time slot, is essential for reducing the operational costs of the C-RAN. Moreover, due to the centralized operation scheme at the BBU, backhaul and processing costs can not be ignored in this setup. Similar to existing works [4],[9] we adopt cloud computing technologies like virtual machines (VMs) to model the processing costs at the cloud. Hence, we follow a user-centric elastic service approach. Each user could be assigned as many VMs as needed to meet the varying user data rate demand. Then we model the processing costs of a specific user as a function of computation capacity referred to as $\mu$, representing the total VM's assigned to user $k$. Thus, in C-RAN the power consumption can be clearly classified into three main parts.
\begin{itemize}
 \item[1-)] The power required for processing the user's data traffic as a function of computation capacity $\mu$ in bit/s.
\vspace{1mm} 
 \item[2-)] The power dissipation due to transmission of data from the cloud to the RRHs through backhaul links. 
\vspace{1mm} 
 \item[3-)] Transmission power for transmitting data through wireless medium between the users and the RRHs.
\end{itemize}
In this work, we investigate the problem of minimizing the power consumption in a green heterogeneous C-RAN (H-CRAN). Such a green H-CRAN consists of multiple macro/micro RRHs of different transmission power capabilities. Each RRH is equipped with a renewable energy source (RES), which can be a solar panel or/and a wind turbine. We also assume that the cloud center is equipped with a large set of RESs compared to each single RRH. In addition, each RRH and the cloud center is provided with a smart meter to control the energy trading process between H-CRAN and the main grid. 

\section{System Model}
Consider a H-CRAN powered with a smart grid in a downlink transmission setup. A heterogeneous set $\mathcal{N} = \left\lbrace 1,2,.., N\right\rbrace $ of multi-antenna RRH's (Macro/Micro), each equipped with $L \geq 1$ antenna is assumed to serve a set $\mathcal{K}$ of single antenna mobile users, where $\mathcal{K} = \left\lbrace 1,2,.., K\right\rbrace $.
We adopt a block-based transmission model in which quasi-static models for both wireless channels and renewable energy process are considered. Thus, the channel coefficients as well as the harvested energy values $P_n^{h}, \; n \in \mathcal{N}$ remain constant during one transmission block and may change from one block to another. 
To facilitate the analysis, we normalize the block duration to unity so that we can use the terms energy and power interchangeably [7], [10]. 
Let the aggregate channel vector of the $k^{th}$ user be $\bold{h}_k = \left[\bold{h}_{k,1}^{T}, \bold{h}_{k,2}^{T},...,\bold{h}_{k,N}^{T}\right]^{T}  \in \mathbb{C}^{LN \times 1}$, where $\bold{h}_{k,n} \in \mathbb{C}^{L \times 1} $ denotes the frequency-flat channel vector between the $n^{th}$ RRH and user $k$. We can write the received signal $y_k \in \mathbb{C}$ at the mobile user $k \in \mathcal{K}$ as
\begin{equation}
y_k = \bold{h}_{k}^{H} \bold{w}_k x_k + \sum\nolimits_{j = 1, j \neq k}^{K} \bold{h}_{k}^{H} \bold{w}_j x_j + n_k,
\end{equation}
where $\bold{w}_k = \left[\bold{w}_{k,1}^{T}, \bold{w}_{k,2}^{T},...,\bold{w}_{k,N}^{T}\right]^{T}  \in \mathbb{C}^{LN \times 1} $ denotes the aggregate beamforming vector of user $k$, $n_k$ refers to the additive white Gaussian noise (AWGN) at the $k^{th}$ receiver and
$x_k \in \mathbb{C}$ is the normalized data symbol with unit power, nominated for the $k^{th}$ user and assumed to be statistically independent from noise, as well as from other users data. We also assume that all the RRHs are synchronized and $x_k$ can be delivered to each RRH via high speed, law latency fiber connections from the cloud.
Although densification of RRH's in H-CRAN can increase the throughput of the network significantly, it will also impose extra charge through backhauling and transmission costs. Hence, we assume in this work that the cloud can decide to turn off any subset of RRH's while ensuring the network is satisfying some QoS requirements. Thus the topology of the network can dynamically be changed to fulfill all the constraints while at the same time minimizing the power consumption. We consider a single user detection strategy at each receiver by treating the interference terms of other users as noise. Thus, one can write the signal to interference ratio (SINR) for each user as a function of beamforming vectors as
\begin{equation}
\mathsf{SINR_{k}} = \small \frac{\left|\bold{h}_{k}^{H} \bold{w}_{k} \right|^{2} }{\sigma_{k}^{2} + \sum\limits\nolimits_{j = 1, j \neq k}^{K} \left|\small \bold{h}_{k}^{H} \bold{w}_{j}\right|^{2} },
\end{equation} 
where $\sigma_{k}^{2}$ is the noise power at the $k^{th}$ user.

\subsection{Modeling C-RAN as a Queue system}
The delay in the C-RAN is caused by two components. The first one due to the processing delay at the cloud center, which is dominated mainly by the  computation capacity $\mu_k$. The second contribution occurs during the transmission phase, which is dominated by the maximum achievable rate of the $k^{th}$ user denoted as $r_{k}$. In order to characterize the overall delay in C-RAN, we adopt a queuing model. We assume the data packets of the $k^{th}$ user are arriving in a random fashion following a poisson distribution with an arrival rate of $\lambda_k, \; k \in \mathcal{K}$. Further, we assume that serving times at the cloud center are exponentially distributed with a mean service time equal to $(1/ \mu_{k})$. To this end, the delay at the \textit{processing} stage can be well described by modeling it with an M/M/1 queue. Thus, the processing delay is simply defined as $\tau_{p_k} = \frac{1}{\mu_{k}-\lambda_{k}} [4] \cdot$ The transmission delay at the RRHs can be also described through modeling the \textit{transmission} stage, again, with an M/M/1 queue whose service node is the achievable rate $r_{k}$ of each user. The two queues can be modeled as two queues in series. This assumption is valid since the transmission rate at the wireless channel is independent from the computational capacity. Hence, at the transmission stage the time delay is given by $\tau_{d_k} = \frac{1}{r_{k}-\lambda_{k}}$. Hence, the total time delay of the $k^{th}$ user in the C-RAN is given as 
\begin{equation}
 T_k(\mu_{k}, r_{k}) = \tau_{d_k} + \tau_{p_k} = \frac{1}{\mu_{k}-\lambda_{k}} + \frac{1}{r_{k}-\lambda_{k}} \cdot
\end{equation}
\subsection{Energy Trading model}
As mentioned before, the smart meter at each RRH and at the BBU is responsible for energy trading between the C-RAN and the main grid. 
So the surplus power at the $n^{th}$ RRH defined as $\max(P_n^{h} - P_n,\; 0)$ can be directly sold to the main grid if the harvested power $P_n^{h}$ is larger than the needed power for transmission, i.e., $P_n$. Otherwise, extra power must be purchased from the main grid, referred to as deficit power and is given by $\max(P_n - P_n^{h},\; 0)$. It is obvious from its expressions that either surplus or deficit power can be strictly positive at one time. We define the cost function at each RRH in terms of power consumption as [7]
\begin{equation}
G(P_n) = \alpha_b \max(P_n - P_n^{h} ,0) - \alpha_s \max(P_n^{h} - P_n,0).
\end{equation} 
Similarly, we define the cost function at the BBU as
\begin{equation}
G(P_e) = \alpha_b \max(P_e - P_{e}^{h} ,0) - \alpha_s \max(P_{e}^{h} - P_e,0),
\end{equation} 
where $\alpha_b$, $\alpha_s$ are the prices of a power unit purchased/sold from/to the main grid, respectively. We assume that $\alpha_s < \alpha_b$ to guarantee a fair power trading between the cloud and the main grid.
$P_e$ and $P_{e}^{h}$ are the power consumption and the power accumulated at the cloud center, respectively.   
\subsection{Problem Formulation}
Our aim is to minimize the power consumption through the entire C-RAN, while satisfying user specific QoS of higher layers (the total delay $T_k$) with respect to some physical layer parameters (rate, transmission power, computation capacity) in a cross-layer optimization scheme. 
This can be done by minimizing the total cost function at the cloud center (in terms of computation capacity and backhaul costs) as well as at the RRHs (in terms of transmission power), while ensuring that end-to-end latency is below a certain threshold. The cost function at RRHs at the cloud center is defined as 
\begin{equation}
\sum\nolimits_{n \in \mathcal{N}} G(P_n)
\end{equation} 
and $G(P_e)$, respectively, where $G(P_n)$, $G(P_e)$ are defined in (4), (5), respectively. 
We assume that $P_{e}^{h} \gg P_n^{h}, \forall n \in \mathcal{N}$, since the cloud center is equipped with a large number of green energy harvesting utilities compared to each single RRH. 
\subsection{Power Consumption at the BBU and RRH clustering}
We model the elastic computational capacity of a cluster of VM's assigned to the $k^{th}$ user as $\mu_k$. Thus, the processing cost of the $k^{th}$ user's data is given as $\varphi(\mu_k) = k_c \mu_{k}^3$ where $k_c > 0$ is a constant. This model has been widely adopted in literature to model the cost aware scalable computation capacity of a cluster of VM's in the cloud computing center.
After processing the data at the cloud, it is forwarded to the RRH's via backhaul high capacity, low-latency fiber links. Since in C-RAN most of processing tasks take place at the cloud center, backhaul costs can not be ignored as in conventional systems. We model this dissipated power as a constant value, $P_C$, associated with data sent to a specific RRH from the cloud.               
Thus, the total cost function at the cloud center in this case can be expressed in terms of power consumption as
$$P_e(\bm{\mu},\mathbf{b}) = k_c \sum\nolimits_{k=1}^{K} \mu_{k}^3 + \sum\nolimits_{n=1}^{N} b_n P_C, $$   
where $\bm{\mu} = \left[\mu_1,\mu_2,\dots,\mu_K \right]^{T}$ and $\mathbf{b} = \left[b_1,b_2,\dots,b_N \right]^{T} \in \left\lbrace 0, 1 \right\rbrace^{N}$  is an \textit{N}-dimensional binary vector used to schedule the RRHs for transmission and to optimize the operation of the C-RAN. Here, $b_n = 0$ means the $n^{th}$ RRH is turned off and it does not participate in the transmission, while $b_n = 1$ means the opposite [5]. Thus, we save energy by turning  off a set of RRHs and let only the ones who participate efficiently in transmission on. This reduces the C-RAN operational costs while the QoS requirements are met. 
\subsubsection{Optimization Problem}
With the cost functions defined in (4), (5), we introduce the following optimization problem to minimize the power consumption over the C-RAN 
\begin{subequations}
\begin{align}
&\underset{ \mathbf{p}_{n}, \mathbf{r}, \bm{\mu}, \mathbf{w}, \mathbf{b}, \mathbf{a}, P_e}{\text{minimize}} 
\begin{aligned}[t]
& \quad \sum\nolimits_{n = 1}^{N} G(P_n) + G\left( P_e\right)\\
\end{aligned} \notag \\
& \text{s.t} \quad  P_n =  \Lambda_n \sum\nolimits_{k = 1}^{K} \norm{\mathbf{w}_{n,k}}_{2}^{2} \\
& P_e = k_c \sum_{k=1}^{K} \mu_{k}^3 + \sum_{n=1}^{N} b_n P_C , \quad \; \mu_{k} \geq \lambda_{k}, r_{k} \geq  \lambda_{k} \\
& r_{k} \leq  B \log(1 + \mathsf{SINR_k}), \quad \; T_k(\mu_k, r_k) \leq \tau_k \\
& \Lambda_n \sum\limits_{k = 1}^{K} \norm{\mathbf{w}_{n,k}}_{2}^{2} \leq  b_{n} P_{n}^{M}, \quad \;  \norm{\mathbf{w}_{n,k}}_{2}^{2} \leq   a_{n,k} P_{n}^{M} \\
& a_{n,k} \leq b_n, \quad \sum\nolimits_{n = 1}^{N} b_n \geq 1 \quad  \text{and} \quad a_{n,k},\;b_n \in \left\lbrace 0, 1\right\rbrace, 
\end{align}
\end{subequations}
where $k \in \mathcal{K}, \; n \in \mathcal{N}, \mathbf{p}_n = \left[P_1,\ldots, P_N\right]^{T}, \mathbf{r} = \left[r_1,\ldots, r_K\right]^{T}$. Before going further into the suggested solution of the above problem, we elaborate briefly on it and the meaning of its physical constraints. Constraints (7a) and (7b) represent the power consumption at the RRHs and the cloud center, respectively.
While (7c) represents a cross-layer constraint, in which we ensure that the total delay $T_k(\mu_k, r_k)$, in (3) is not exceeding a given QoS value $\tau_k$.
The scheduling variables $a_{n,k}$ associate the $k^{th}$ user with the $n^{th}$ RRH, such that if $a_{n,k} = 0$, the user $k$ is not scheduled to be served by the RRH $n$ and therefore, the beamforming vector $\mathbf{w}_{n,k}$ is a null vector of size $L$, i.e, $\mathbf{w}_{n,k} = \mathbf{0}_L \hspace{2mm} \text{when} \hspace{2mm} a_{n,k} = 0, \quad \forall k \in \mathcal{K}, \forall n \in \mathcal{N}$. Constraint (7e) ensures that users would be scheduled to a RRH only if it is on (e.g $b_n = 1$) and at least one RRH is serving the users within the cloud network. The constant $\frac{1}{\Lambda_n}$ represents here the power amplifier efficiency of RRH $n$, and $P_{n}^{M}$ is the maximum transmission power assigned to the $n^{th}$ RRH.
The problem (7) is challenging to solve due to the following reasons,
\begin{enumerate}
  \item The problem belongs to a  mixed integer non linear program (MINLP) class, which is known to be a NP-hard problem. 
  \item Even if we relax the integer constraints (7e) to be continuous ones; the resulting non linear program would itself be a non-convex program.  
  \item The parameter space of the problem can grow very large as the number of users or clusters connected to the cloud increases, which make it very difficult to solve efficiently in reasonable time.
\end{enumerate}
The first step to solve the problem (7) is to convexify the constraints (7a)-(7c) and the objective function. By doing this the program resulting from relaxing the integer constraint (7e) is a convex one. Objective function is convexified by rewriting (4) as 
  \begin{equation*}
    G^{'}(P_n) =  \psi \left|P_n - P_n^{h} \right| + \phi \left(P_n - P_n^{h}\right),
  \end{equation*}
where 
  \begin{equation*}
 \psi = \frac{\alpha_b - \alpha_s}{2}, \phi = \frac{\alpha_b + \alpha_s}{2}
  \end{equation*}
we get similar expression for cost function at the cloud center $G^{'}(P_e)$ by rewriting (5) in the same manner.   
To convexify (7c), we make use of the relation between the maximum rate on a wireless link and the mean square error (MSE) [2], as given by the following lemma \\
\textit{Lemma 1}: For a wireless link, the maximum achievable rate in (7c), can be equivalently written in terms of MSE as
\begin{equation}
r_k =  \underset{u_k,v_k}{\text{max}} \left[1 + log(v_k) - v_k e_k(\mathbf{w},u_k) \right], 
\end{equation}
where $v_k$ is a weighting coefficient of the MSE, and $u_k \in \mathbb{C}$ is the linear receiver coefficient, applied by the single antenna user to decode the desired signal from (1). Here, 
$e_k(\mathbf{w},u_k)$ is the MSE given by
\begin{equation}
e_k(\mathbf{w},u_k) = E\left\lbrace (x_k - \hat{x_k})(x_k - \hat{x_k})^H\right\rbrace, 
\end{equation} 
where $E\left\lbrace \cdot \right\rbrace $ is the statistical expectation operator, $\hat{x_k} = u_{k}^{H} y_k$ is the decoded symbol at the receiver and $y_k$ is given in (1). 
\normalsize
Now, we can write the reformulated problem (7) as
\begin{subequations}
\begin{align}
&\underset{\mathbf{u}, \mathbf{v},\mathbf{p}_n, \mathbf{r},\bm{\mu},\mathbf{w},\mathbf{b},\mathbf{a}, P_e}{\text{minimize}} 
\begin{aligned}[t]
& \quad \sum\nolimits_{n = 1}^{N} G^{'}(P_n) + G^{'}\left( P_e\right) \\
\end{aligned} \notag \\
& \text{s.t} \quad  (7a), (7b), (7d), (7e) \\
& r_k \leq r_k^{M} \quad T_k(\mu_k, r_k) \leq \tau_k, 
\end{align}
\end{subequations}
where  
\begin{equation}
r_k^{M} =  B \Big(  c_{1,k} +\mathbf{c}_{2,k}\mathbf{w}_k + \mathbf{w}_{k}^{H}\mathbf{c}_{3,k} - c_{4,k} \small \sum\nolimits_{j \in \mathcal{K}} \left|\mathbf{h}_{k}^{H} \mathbf{w}_j \right|^2 \Big)
\end{equation} 
\begin{equation*}
c_{1,k} = 1+\log_2(v_{k}^{*})-v_{k}^{*}\big(1 + \sigma_{k}^2 \left| u_{k}^{*}\right| ^2 \big), \; \mathbf{c}_{2,k} =  2 v_{k}^{*} u_{k}^{*}\mathbf{h}_{k},
\end{equation*} 
$\mathbf{c}_{3,k} = \mathbf{c}_{2,k}^{H}$ and $c_{4,k} = v_{k}^{*}  \left| u_{k}^{*}\right| ^2$. Here, $v_{k}^{*}, u_{k}^{*}$ represent the optimal solution to the optimization problem in (8), and can be expressed in a closed form expressions as
 \begin{subequations}
   \begin{alignat}{2}
     u_{k}^{*}  & = \frac{\mathbf{h}_{k}^{H} \mathbf{w}_k}{\sigma_{k}^2 + \sum\limits_{j \in \mathcal{K}} \left|\mathbf{h}_{k}^{H} \mathbf{w}_j \right|^2}  \\
     v_{k}^{*}  & = \frac{1}{e_k(\mathbf{w},u_{k}^{*})} = \frac{1}{1- \mathbf{w}_{k}^{H}\mathbf{h}_{k} u_{k}^{*}} \cdot      
   \end{alignat}
   \end{subequations}  
Although problem (10) include two additional optimization variables $(\mathbf{u} = \left[u_1, u_2,\ldots,u_K \right]^{T}, \mathbf{v} = \left[v_1, v_2,\ldots,v_K \right]^{T})$, it is easier to solve than directly solving problem (7). 
Similar to [3], we suggest the following approach to solve problem (10), by splitting it into two problems. The first one is solved by using the weighted minimum mean square error (WMMSE) algorithm to update the coefficients $\mathbf{u}, \mathbf{v}$. This is done by using formulas (12a, 12b).
The outline of this algorithm is given in table 1. After each iteration $i$, we plug in the updated coefficients $(\textbf{u}^{i}, \textbf{v}^{i})$ in the constraint (10b) to update the upper bound on the maximum achievable rate $(r_{k,i}^{M})$ using the equation (11). Thus, we get the following MINLP in the $i^{th}$ iteration of WMMSE,
\begin{subequations}
\begin{align}
&\underset{\mathbf{p}_n, \mathbf{r}, \bm{\mu}, \mathbf{w}, \mathbf{b}, \mathbf{a}, P_e}{\text{minimize}} 
\begin{aligned}[t]
& \quad \sum\nolimits_{n = 1}^{N} G^{'}(P_n) + G^{'}\left( P_e\right) \\
\end{aligned} \notag \\
& \text{s.t} \quad  (7a), (7b), (7d), (7e) \\
& r_k \leq r_{k,i}^{M} \quad T_k \leq \tau_k. 
\end{align}
\end{subequations}
By looking carefully at problem (13), we notice that the constraint related to the rate $r_k$ in (13b) complicates matters, since it links all the optimization variables. Thus, the problem without this constraint can be easily separated into two problems defined on two group of variables. The first one which contains only $(\mathbf{P}_n, P_e,\mathbf{r},\bm{\mu})$, while the second one contains $(\mathbf{w},\mathbf{b},\mathbf{a})$. To make this separation between two set of variables possible and to linearize the equality constraints in (7a) and (7b), we introduce new variables $ \mu_{k}^{'} = \mu_{k}^{3} \hspace{2mm} \text{and} \hspace{2mm} \mathbf{t} = \left[\mathbf{t}_{1}^{T}, \mathbf{t}_{2}^{T},\ldots,\mathbf{t}_{N}^{T} \right]^{T}$ where  $\mathbf{t}_n = \left[t_{n,1}, t_{n,2},\ldots,t_{n,K} \right]^{T} \in \mathbb{R_{+}}^{K \times 1}$. Similarly to $\mathbf{t}$, we define the variable $\mathbf{x} = \left[\mathbf{x}_{1}^{T}, \mathbf{x}_{2}^{T},\ldots,\mathbf{x}_{N}^{T} \right]^{T} \in \mathbb{R_{+}}^{NK \times 1}$ and reformulate the problem (13) accordingly.
\begin{subequations}
\begin{align}
&\underset{\mathbf{p}_n, \mathbf{r}, \bm{\mu^{'}}, \mathbf{w}, \mathbf{b}, \mathbf{a}, P_e, \mathbf{x}, \mathbf{t}}{\text{minimize}} 
\begin{aligned}[t]
& \quad \sum\nolimits_{n = 1}^{N} G^{'}(P_n) + G^{'}\left( P_e\right) \\
\end{aligned} \notag \\
& \text{s.t}  \quad (7e), (13b), \; \quad \text{with} \quad T_k\Big(\sqrt[3]{\mu_{k}^{'}},r_k\Big) \leq \tau_k \\ 
& P_e = k_c \sum\nolimits_{k=1}^{K} \mu_{k}^{'} + \sum\limits_{n=1}^{N} b_n P_C \\ %
& P_n =  \Lambda_n \mathbf{d}_{n}^{T} \mathbf{x}, \quad \;  \mathbf{t} = \mathbf{x} \\
& \norm{\mathbf{w}_{n,k}}_{2}^{2} \leq \mathbf{d}_{n,k}^{T} \mathbf{t} , \quad \; \sqrt[3]{\mu_{k}^{'}} \geq \lambda_{k},\; r_{k} \geq  \lambda_{k} \\
&\Lambda_n \mathbf{d}_{n}^{T} \mathbf{x}  \leq b_n \cdot P_{n}^{M},  \quad \;  \mathbf{d}_{n,k}^{T} \mathbf{x} \leq a_{n,k}\cdot P_{n}^{M},
\end{align}
\end{subequations}
\vspace{-2mm}
where,
\small
\begin{align*}
\mathbf{d}_n &     = [\mathbf{0}_{(n-1)K},\mathbf{1}_{K}, \mathbf{0}_{(N-n)K}]^{T} \\
\mathbf{d}_{n,k} & = [\mathbf{0}_{(n-1)K},\mathbf{0}_{(k-1)}, 1, \mathbf{0}_{K-k},\mathbf{0}_{(N-n)K}]^{T},
\end{align*}
where $\mathbf{1}_{K}$ is a vector of length $ K $ and all its elements are equal to one.   
\normalsize
Note that with this notation, both vectors $\mathbf{x}, \mathbf{t}$ represent nothing but the power assigned to each user at each RRH terminal. 
\vspace{-3mm}
\subsection{ADMM Algorithm}
As problem (14) is now amenable for a distributed solution, we adopt the ADMM algorithm which can be compatibly written as [6]  
\begin{align}
\underset{\mathbf{y},\mathbf{z}}{\text{minimize}} \quad f(\mathbf{y}) + g(\mathbf{z}) \nonumber \\
\text{subject to}  \quad \mathbf{A}\mathbf{y}_m + \mathbf{B}\mathbf{z} = 0  \nonumber\\
 \quad  \mathbf{y} \in \mathcal{C}_1 , \; \mathbf{z} \in \mathcal{C}_2,  
\end{align}
where
\vspace{-2mm}
\begin{align*}
\mathbf{y}_m  &= \left[\mathbf{p}_{n}^{T}, P_e, \bm{\mu}^{T}; \mathbf{t}^{T}\right]^{T}, \quad \;  \mathbf{z} =  \left[\mathbf{b}^{T}; \mathbf{a}^{T}; \mathbf{x}^{T}\right]^{T},\\
\mathbf{y}    &= \left[\mathbf{y}_{m}^{T}; \mathbf{r}^{T}; \mathbf{w}^{T}\right]^{T} \in \mathbb{R}^{((N+1)(K+1)+K(1+NL))\times 1} 
\end{align*}
\vspace{-3mm}
and
\begin{align*}
\mathbf{A} \in \mathbb{R}^{(N(K+1)+1)\times (N+1)(K+1)}, \;  \mathbf{B} \in \mathbb{R}^{(N(K+1)+1)\times(2K+1)N}\cdot
\end{align*}
Here, $\mathcal{C}_2$ is a set defined by mixed integer constraints given by (7e) and (14e), while $\mathcal{C}_1$ is a convex set defined by remaining group of constraints in (14), apart from equality constraints. The matrices $\mathbf{A}, \mathbf{B}$ are defined in accordance with equality constraints of problem (14), i.e., (14b) and (14c). $$f(\mathbf{y}) = f(\mathbf{p}_n, P_e) = \sum\limits_{n = 1}^{N} G^{'}(P_n) + G^{'}(P_e),$$
while $g(\mathbf{z})$ is the indicator function to the set $\mathcal{C}_2$. First we define the augmented Lagrangian function as,
\begin{alignat}{2}
L_{\rho}\left(\mathbf{y}, \mathbf{z}, \bm{\gamma} \right) = f(\mathbf{y}) + g(\mathbf{z}) + \bm{\gamma}^{T}\left(\mathbf{A}\mathbf{y}_m + \mathbf{B}\mathbf{z} \right) \nonumber \\ 
 + \left(\frac{\rho}{2}\right) \norm{\mathbf{A}\mathbf{y}_m + \mathbf{B}\mathbf{z}}_{2}^{2} .
\end{alignat}
Where $\rho > 0$ is penalty parameter, and $\bm{\gamma} \in \mathbb{R}^{(N(K+1)+1)\times 1}$ are the Lagrange multipliers.
Then, the ADMM consists of the following distributed iterations,
\begin{align}
\mathbf{y}^{k+1}       & = \underset{\mathbf{y} \in \mathcal{C}_1}{\text{argmin}}\; L_{\rho}\left(\mathbf{y}, \mathbf{z}^{k}, \bm{\gamma}^{k} \right) \\
\mathbf{z}^{k+1}       & = \underset{\mathbf{z} \in \mathcal{C}_2}{\text{argmin}}\; L_{\rho}\left(\mathbf{y}^{k+1}, \mathbf{z}, \bm{\gamma}^{k} \right) \\
\bm{\gamma}^{k+1}      & = \bm{\gamma}^{k} + \rho\left(\mathbf{A}\mathbf{y}_{m}^{k+1} + \mathbf{B}\mathbf{z}^{k+1} \right). 
\end{align}
Subproblem (17) is a convex optimization problem and can be solved efficiently. Subproblem (18) is a MINLP one (the objective is quadratic, however the constraints are linear), but it is simplified compared to the original one (since the later contains quadratic constraints). 
The last step in ADMM is a central step, where we gather the updated primal variables of steps (17) and (18) to update the Lagrange multipliers in (19). \\
\begin{table}
\caption{WMMSE algorithm embedded with ADMM iterations}
\noindent\fbox{%
\begin{minipage}{\dimexpr\linewidth-2\fboxsep-2\fboxrule\relax}
\begin{enumerate}
\item[1]\textbf{Initialization}, i= 0. Initialize $\textbf{u}^{0}$, $\textbf{v}^{0}$ using a feasible initial beamforming vector $\mathbf{w}^{0}$ and equation (12).
\item[2] \textbf{Repeat}
\item[3] $\textbf{u}^{i}$ is updated using (12a).
\item[4] $\textbf{v}^{i}$ is updated using (12b).
\item[5] All the remaining variables are updated using ADMM algorithm to solve problem (15), using iterations (17) to (19) as explained in section $E$. 
\item[6] $i \leftarrow i+1$
\item[7] $\textbf{Until}$ some suitable convergence criterion is met.
\end{enumerate}
\end{minipage}%
} 
\end{table}
\section{Numerical simulation}
In this section we consider a H-CRAN system consists of a high power RRH, with six low power helping RRH nodes distributed on a circle of radios 0.6 km around it. Each RRH is provided with a solar panel and/or a wind turbine. We assume non-uniform distribution of RES at base stations, which means that the harvested energy value at each RRH can and will be different from other RRHs. The cloud center, though, is equipped with a high number of RES. We use the channel pathloss model in which the channel coefficients between the $l^{th}$ antenna of the $n^{th}$ RRH and the $k^{th}$ user are given by [5],
\begin{equation*}
h_{nk}^{l} = \Gamma_{nk}^{l} \sqrt{G\beta d_{nk}^{-\alpha}\delta_{nk}}, \; n \in \mathcal{N}, \; k \in \mathcal{K}, \; l =1,\ldots, L,
\end{equation*} 
where $\Gamma_{nk}^{l} \sim \mathcal{CN}(0,1)$ represents the small scale fading, $G$ models the antenna power gain and is chosen to be $9 dB$ and $\delta_{nk}$ models the shadowing effect with $\delta_{dB} \sim \mathcal{CN}(1, 6.31)$. The term $\beta d_{nk}^{-\alpha}$ models the pathloss, which uses the 3GPP specifications, given as
\begin{equation}
p_{nk}(dB) = 128.1 + 37.6 \log_{10} (d_{nk}).
\end{equation}  
Here, $d_{nk}$ is the distance in km.
The harvested energy values $\left\lbrace P_n^{h} \right\rbrace_{i=1}^{N} $ are based on a real world measurements of both solar and wind energy production, [8]. Finally, we consider that the network sells surplus energy to the grid with the price $\alpha_s = 0.1 \alpha_b$, where $\alpha_b =  1 \text{Unit}/ \text{watts}$ is the price of energy purchased from the grid.
we look the behavior of power consumption cost at the green C-RAN, as we increase the average data arrival rate of users. We assume a time delay QoS value of $\tau = 1 ms $. We further assume that each RRH is provided with $L = 4$ antennas serving 4 users, under this setup we get the Fig. 1 which shows the behavior of optimal power consumption cost in currency unit/ watts as a function of the incoming data rate ($\lambda_k$) in Mbps. According to the Fig. 1, there is a balance point at 6.15 Mbps. At this point the cost function is zero which means that C-RAN uses all the harvested energy to satisfy the user requirements without the need to buy energy from the main grid, but also without selling any. If the users requirements were below the data rate at the balance point, the C-RAN can sell surplus power to the main grid, hence, the cost function is negative, otherwise C-RAN must buy deficit power to cover its needs. We also note different slopes of the optimal cost function at the two regions around the balance point, this is due to the difference in selling/buying prices ($\alpha_s, \alpha_b$), in this work we assumed those prices to be given a priori, for future works it would be interesting to show the influence of dynamic pricing on the energy trading system. We also examine the convergence behavior of the ADMM algorithm in this medium-size network. Fig. 2 shows the convergence of the objective function for this setup in one run of step 5 in table 1. Here, we use the primal residual $r$ and dual residual $s$ as stopping criterion for the algorithm to guarantee a good quality solution of the problem [6]. 
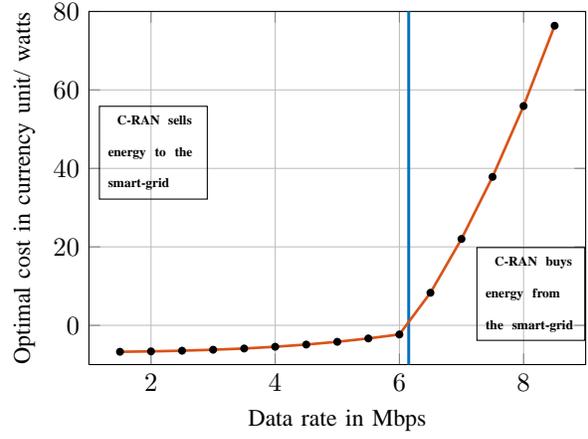
\begin{figure}[h]
%
%
\definecolor{mycolor1}{rgb}{0.00000,0.44700,0.74100}%
\definecolor{mycolor2}{rgb}{0.85000,0.32500,0.09800}%
\begin{tikzpicture}

\begin{axis}[%
width=2.6in,
height=1.85in,
at={(0.1in,12.224in)},
scale only axis,
xmin=1,
xmax=9,
xlabel={\small Data rate in Mbps},
xmajorgrids,
ymin=-10,
ymax=80,
ylabel={\small Optimal cost in currency unit/ watts},
ymajorgrids,
axis background/.style={fill=white}
]
\addplot [color=mycolor1,solid,forget plot]
  table[row sep=crcr]{%
1.5	-6.73579869509074\\
2	-6.63049636422253\\
2.5	-6.46203712037096\\
3	-6.21497117708362\\
3.5	-5.89771053325484\\
4	-5.45524172392123\\
4.5	-4.89272173627559\\
5	-4.1898937266899\\
5.5	-3.33651848931755\\
6	-2.31759328823891\\
6.5	8.3085462181868\\
7	22.0200356312039\\
7.5	37.8315242791545\\
8	55.8883289053625\\
8.5	76.3563524684376\\
};
\addplot [color=mycolor1,solid,line width=1.1pt,forget plot]
  table[row sep=crcr]{%
6.15	-10\\
6.15	80\\
};
\draw [] (rel axis cs:0.02,0.6) node[right,draw,text width=1.2cm, align=left] {\tiny \textbf{ C-RAN sells energy to the smart-grid}};


\draw [] (rel axis cs:0.78,0.2) node[right,draw,text width=1.2cm, align=left] {\tiny \textbf{ C-RAN buys energy from the smart-grid}};
\addplot [color=mycolor2,solid,line width=1.0pt,mark size=1.0pt,mark=*,mark options={solid,fill=black,draw=black},forget plot]
  table[row sep=crcr]{%
1.5	-6.73579869509074\\
2	-6.63049636422253\\
2.5	-6.46203712037096\\
3	-6.21497117708362\\
3.5	-5.89771053325484\\
4	-5.45524172392123\\
4.5	-4.89272173627559\\
5	-4.1898937266899\\
5.5	-3.33651848931755\\
6	-2.31759328823891\\
6.5	8.3085462181868\\
7	22.0200356312039\\
7.5	37.8315242791545\\
8	55.8883289053625\\
8.5	76.3563524684376\\
};
\end{axis}
\end{tikzpicture}%
\caption{Operational costs of the C-RAN as a function of data rate}
\end{figure}
\begin{figure}[ht]
\input{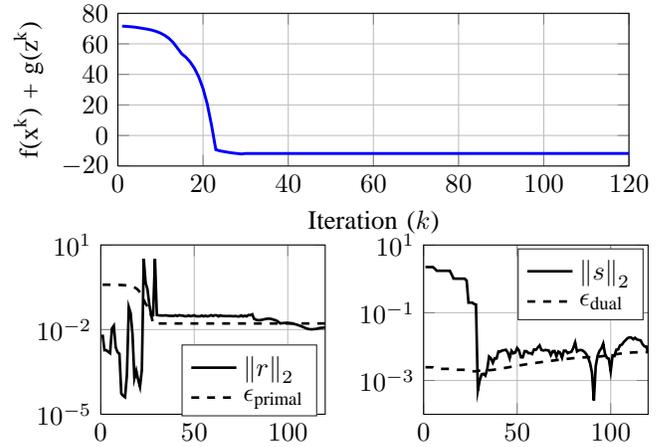}   
\caption{One run of the ADMM algorithm. The first figure shows the objective progress from the initial vector (all RRHs are on and transmit with full power), in the two sub figures below we show the primal residual to the left and dual one to the right. They determine the stopping criterion of ADMM.}
\end{figure}

\end{document}